\documentclass[12pt]{article}
\usepackage{graphicx}
\usepackage{natbib} 
\usepackage{url} 

\usepackage{amsmath}
\usepackage{amssymb}
\usepackage{amsfonts}
\usepackage{multirow}
\usepackage{amsthm}

\usepackage{moreverb,url}

\usepackage{graphicx}
\usepackage{amsmath}
\usepackage{times}
\usepackage{bm}
\usepackage{enumerate}
\usepackage{mathtools}
\usepackage{setspace}
\usepackage{array,booktabs}
\usepackage{tikz}
\usepackage{natbib}
\usepackage{mathtools}
\usepackage{chngcntr}
\usepackage{algorithm}
\usepackage[utf8]{inputenc}
\usepackage{xcolor, soul}
\usepackage{hyperref}

\DeclareMathOperator*{\argmax}{arg\,max}
\DeclareMathOperator*{\argmin}{arg\,min}

\theoremstyle{definition}

\newcommand{\blind}{1}

\begin{document}

\def\spacingset#1{\renewcommand{\baselinestretch}%
{#1}\small\normalsize} \spacingset{1}

\date{}

\if1\blind
{
	\title{\bf Optimizing Graphical Procedures for Multiplicity Control in a Confirmatory Clinical Trial via Deep Learning}
 \author{Tianyu Zhan\thanks{Data and Statistical Sciences, AbbVie Inc., 1 North Waukegan Road, North Chicago, IL 60064. Email of the corresponding author: tianyu.zhan@abbvie.com}, 
 	Alan Hartford\thanks{Statistical and Quantitative Sciences, Data Sciences Institute, Research and Development, Takeda Pharmaceuticals USA, Inc., Cambridge, MA 02139},
 	Jian Kang\thanks{Department of Biostatistics, University of Michigan, Ann Arbor, MI 48109},
 	and Walter Offen\thanks{Retired from AbbVie Inc.}}
	\maketitle
} \fi

\if0\blind
{
	\title{\bf Deep Neural Networks Guided Ensemble Learning for Point Estimation}
	\maketitle
} \fi

\bigskip
\begin{abstract}
In confirmatory clinical trials, it has been proposed to use a simple iterative graphical approach to construct and perform intersection hypotheses tests with a weighted Bonferroni-type procedure to control type I errors in the strong sense. Given Phase II study results or other prior knowledge, it is usually of main interest to find the optimal graph that maximizes a certain objective function in a future Phase III study. In this article, we evaluate the performance of two existing derivative-free constrained methods, and further propose a deep learning enhanced optimization framework. Our method numerically approximates the objective function via feedforward neural networks (FNNs) and then performs optimization with available gradient information. It can be constrained so that some features of the testing procedure are held fixed while optimizing over other features. Simulation studies show that our FNN-based approach has a better balance between robustness and time efficiency than some existing derivative-free constrained optimization algorithms. Compared to the traditional stochastic search method, our optimizer has moderate multiplicity adjusted power gain when the number of hypotheses is relatively large. We further apply it to a case study to illustrate how to optimize a multiple testing procedure with respect to a specific study objective.
\end{abstract}

\noindent%
\vfill

\newpage
\spacingset{2} 

\section{Introduction}

Most clinical trials performed in drug development contain multiple endpoints to assess the effects of the drug and to document the ability of the drug to favorably affect one or more disease characteristics \citep{kelly2015short, kazda2016evaluation, guidance2017multiple}. Adequate multiple testing procedures (MTPs) are required to protect the family-wise error rate (FWER), which is the probability of rejecting at least one true null hypothesis. Proper MTPs should be employed to reflect relative importance of multiple endpoints and different study objectives. A variety of weighted Bonferroni-based test procedures have been proposed, for example, the weighted or unweighted Bonferroni-Holm procedure \citep{holm1979simple}, fixed sequence tests \citep{westfall2001optimally}, the fallback procedure \citep{wiens2003fixed} and gatekeeping procedures based on Bonferroni adjustments \citep{dmitrienko2003gatekeeping}.

Those aforementioned approaches usually need to specify a large number of intersection hypotheses tests according to the closure principle \citep{marcus1976closed}. It is often difficult to apply those methods in practice, especially when the number of endpoints is relatively large. Taking a study with $10$ hypotheses as an example, there are $2^{10} - 1 = 1,023$ intersection hypotheses in the full closure. \citet{bretz2009graphical} propose a graphical approach to represent a wide range of MTPs with weighted Bonferroni tests for intersection hypotheses. Based on the monotonicity for local significance levels, the graphical approach essentially establishes a shortcut to the closure test procedure and leads to a sequentially rejective procedure with up to $m$ steps, where $m$ is the number of null hypotheses to be tested. The graphical representation of this approach is easier to communicate with clinical teams and facilitates the discussion of different strategies to fulfill distinct study objectives. However, choosing a graph in complex testing situations can still be overwhelming. While practical considerations for achieving the most desired drug label may take precedence over the most efficient graphical testing procedure, the decision of which graphical testing to choose will be served well by being informed of the optimal graph with respect to an objective function. 

Since graphical approaches analytically control FWER at a desired level in the strong sense \citep{bretz2009graphical}, can we further identify an optimal graph in a confirmatory trial with respect to certain objective functions based on prior knowledge such as from Phase II studies? \cite{rubin2006method} and \cite{wasserman2006weighted} studied the power function for the weighted Bonferroni procedure, but the graphical approach we considered is more general. As can be seen later in this article, it is difficult to evaluate the objective function and its derivatives in closed forms due to the complex correlations between the decision functions from different endpoints. The stochastic search method (SSM; \cite{zabinsky2013stochastic}) is popular in practice due to its ease of implementation. This approximating approach is to find the graph with the maximum working objective function among a certain number of randomly simulated candidates under constraints. However, this method is very likely to miss the optimal target when the number of endpoints is relatively large, as demonstrated later in this article. 

Another stream is to adopt existing derivative-free constrained optimization methods. The ability to handle both bounded and inequality constraints is desired to accommodate different study objectives and the constraints in the graphical approach. There are vast numbers of those approaches available in the field of machine learning, but their performances in finding either global or local optima vary depending on the problem at hand \citep{kramer2011derivative}. In this article, we evaluate the performance of the Improved Stochastic Ranking Evolution Strategy (ISRES; \cite{runarsson2005search}) and the Constrained Optimization BY Linear Approximations (COBYLA; \cite{powell1994direct}) on optimizing graphical approaches by simulation studies. They are readily implemented in the R package {\it{nloptr}} \citep{nloptr}. As an alternative, we propose an optimization framework based on deep learning with moderate power gain and tolerable extra computing time. 

Deep learning has made substantial success in various domains such as image recognition and natural language processing \citep{goodfellow2016deep}, and is also receiving attention from the pharmaceutical industry. For example, \cite{liang2018estimating} propose a novel outcome weighted deep learning algorithm to estimate individualized optimal combination therapy; and \cite{zhan2019targeting} construct test statistics based on Deep Neural Networks to increase power in sample size reassessment adaptive clinical trials. In this article, we utilize feedforward neural networks (FNNs) to approximate the underlying complex objective function and further identify the optima with available gradient information. Our method has several distinguishing features. First of all, flexible utility functions can be defined to accommodate different study objectives. Moreover, our method is able to perform optimization when certain structures in the graph are fixed. Additionally, gradients are readily available from the fitted FNN, and do not need to be computed from the complex objective function, which is offen not feasible even with numerical methods. Compared with the two derivative-free optimization approaches, our FNN-based optimizer offers a better balance between time efficiency and robustness. More details are provided in Section \ref{s:sim}.  

The remainder of this article is organized as follows. In Section 2, we review the graphical approach for multiple hypotheses testing and further define the objective function to optimize. In Section 3, we introduce our optimizing methods via deep learning techniques. Simulations under multiple scenarios are conducted to evaluate the performance of our procedures in Section 4. In Section 5, we implement our method in a case study. Finally, concluding remarks are provided in Section 6. 

\section{The graphical approach to sequentially rejective multiple testing procedures}

In this section, we first review the graphical approach as an MTP which strongly controls the family-wise error rate (FWER) at a nominal level $\alpha$ in Section \ref{s:review}. It is essentially a shortcut to the closed testing procedure with the weighted Bonferroni test for intersection hypotheses. In Section \ref{s:obj}, we introduce an objective function to evaluate the performance of a specific graph. 

\subsection{Review of the graphical approach}
\label{s:review}

Suppose in a clinical trial, we are interested in testing $m$ elementary null hypotheses, $H_1, H_2, ..., H_m$, with observed unadjusted $p$-values $\boldsymbol{p} = (p_1, p_2, ..., p_m)$. Let $\alpha$ denote the one-sided FWER (usually $\alpha = 0.025$ in practice). A multiple testing procedure (MTP) is said to control the FWER at $\alpha$ in the strong sense that the probability of rejecting at least one true null hypothesis does not exceed $\alpha$ under any configuration of true and false null hypotheses. The MTPs can be derived from the closure principle \citep{marcus1976closed}, which requires $2^m-1$ local $\alpha$-level tests of each non-empty intersection hypothesis $H(I) = \cap_{i \in I} H_i$, where $I \subseteq M = \{1, 2, ..., m\}$ \citep{tamhane2018advances}. An intersection hypothesis $H(I)$ is rejected if and only if all $H(J)$ for $J\supseteq I$ are rejected by their $\alpha$-level tests. As a shortcut, if the local tests are consonant \citep{gabriel1969simultaneous}, then the corresponding MTP requires only up to $m$ local tests. For example, the Holm MTP uses Bonferroni tests as the local tests for all intersection hypotheses \citep{holm1979simple}.

The graphical approach defines a shortcut MTP for a closed testing procedure with weighted Bonferroni tests for the intersection hypotheses to strongly control the FWER at $\alpha$. Specifically, the weighted Bonferroni rejects $H(I)$ if $\left\{\min_{i \in I} \left( p_i/w_i\right)\right\} \leq \alpha$, where $\sum_{i=1}^{|I|} w_i =1$ and $|I|$ denotes the number of elements in $I$. In order to specify a graph, one needs to define two components: the initial alpha allocation vector $\boldsymbol{\alpha}$ and the transition matrix $\boldsymbol{T}$. Let $\boldsymbol{\alpha} = (\alpha_1, \alpha_2, ..., \alpha_m)$ denote the initial assignment of overall significance level under the constraint,
\begin{equation}
	\label{equ:cons_alpha}
	\sum_{i=1}^m \alpha_i = \alpha.
\end{equation}
Note that the equality sign in (\ref{equ:cons_alpha}) is to make full use of all available significance levels to gain the highest power. It can be replaced by the sign ``$\leq$'' while still controlling FWER at $\alpha$. The transition matrix $\boldsymbol{T}$ is an $m \times m$ matrix, where each element $T_{ij}$ specifies the proportion of local significance level $\alpha_i$ that is passed to $H_j$ if $H_i$ is rejected at $\alpha_i$. For all $i, j = 1, 2, ..., m$, $T_{ij}$ has to satisfy the following conditions:
\begin{align}
	\label{equ:cons_G}
	& 0 \leq T_{ij} \leq 1, \:\:\:\:\:\: T_{ii} = 0, \:\:\:\:\:\: \sum_{k=1}^m T_{ik} = 1.
\end{align}
We further use $g(\boldsymbol{\alpha}, \boldsymbol{T})$ to denote a graph with vector $\boldsymbol{\alpha}$ and matrix $\boldsymbol{T}$. The graphical approach $g(\boldsymbol{\alpha}, \boldsymbol{T})$ can represent a variety of weighted Bonferroni-based test procedures.

\begin{figure}[h]
	\centering
	\includegraphics[width=0.8\linewidth]{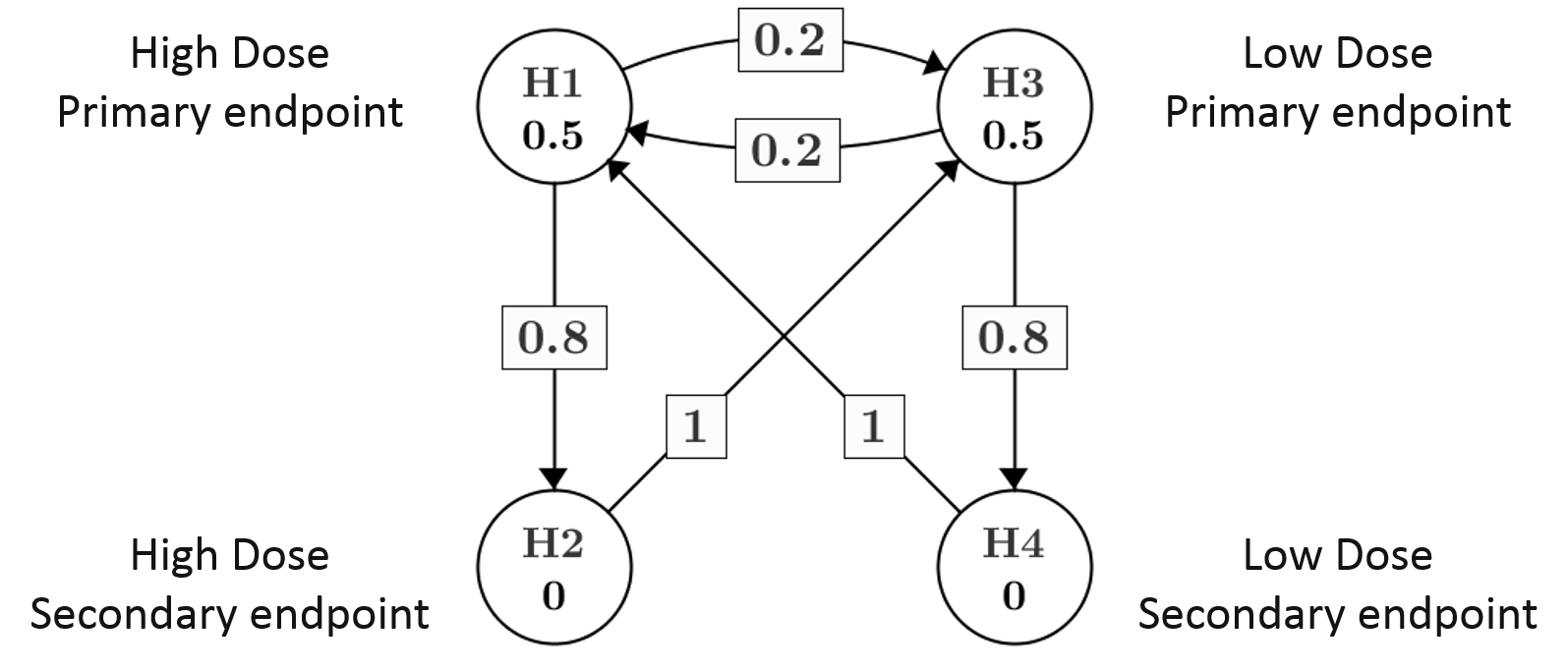}
	\caption{A motivating example of a graphical approach for multiplicity control of two doses and two endpoints.}
	\label{F:motivating}
\end{figure}

Consider a motivating example of a Phase III clinical trial with two doses (high and low) and two endpoints (primary and secondary) in each dose. The team may want to consider a design represented by the graphical procedure in Figure \ref{F:motivating}. One first tests the primary endpoint in each dose with $0.5 \times \alpha$; $80\%$ of it will be passed to the secondary endpoint and $20\%$ to the primary endpoint in the other dose if rejected. Once rejected, the significance level of the secondary endpoint can also be fully recycled to the primary endpoint in the alternative dose. In this case, the initial alpha allocation vector $\boldsymbol{\alpha}$ is $(0.0125, 0, 0.0125, 0)$ and the transition matrix $\boldsymbol{T}$ is given by
\[ 
\boldsymbol{T} = \left (
\begin{tabular}{cccc}
0 & 0.8 & 0.2 & 0 \\
0 & 0 & 1  & 0 \\
0.2 & 0 & 0 & 0.8 \\
1 & 0 & 0 & 0 
\end{tabular}
\right ).
\]


Given the observed unadjusted $p$-value vector $\boldsymbol{p}$, the graphical approach establishes a sequentially rejective test procedure that is illustrated in Algorithm \ref{alg:graph}. Basically, one tests the most significant hypothesis with its non-zero local significance level. If it is rejected, then update the graph according to the pre-specified rules. We further define a decision function $D_i \left(\boldsymbol{\alpha}, \boldsymbol{T}, \boldsymbol{p} \right)$ for endpoint $i$, which takes value $1$ if its null hypothesis is rejected under a graphical approach $g(\boldsymbol{\alpha}, \boldsymbol{T})$, and $0$ otherwise. 
\begin{algorithm}
	\caption{Graphical approaches \citep{bretz2009graphical}}
	\label{alg:graph}
	0. Set $I = M$.\\
	1. Let $j = \text{argmin}_{i \in I} \:\: p_i/\alpha_i$.\\
	2. If $p_j \leq \alpha_j$, then reject $H_j$; otherwise stop.\\
	3. Update the graph:
	\begin{align*}
		& I \rightarrow I \setminus \{j\} \\ 
		& \alpha_l=\left\{
		\begin{array}{ll}
			\alpha_l + \alpha_j T_{jl}, \:\:\:\:\:\:\:\ l \in I\\
			0, \:\:\:\:\:\:\:\:\:\:\:\:\:\:\:\:\:\:\:\:\:\:\:\: \text{otherwise}
		\end{array}
		\right.\\
		& T_{lk} = \left\{
		\begin{array}{ll}
			\dfrac{T_{lk}+T_{lj}T_{jk}}{1-T_{lj}T_{jl}}, \:\:\:\:\:\: l, k \in I, \:\: l \neq k\\
			0, \:\:\:\:\:\:\:\:\:\:\:\:\:\:\:\:\:\:\:\:\:\:\: \text{otherwise}
		\end{array}
		\right.\\
	\end{align*}
	4. If $|I| \geq 1$, go to step 1; otherwise stop. 
\end{algorithm}

Since all graphs under constraints (\ref{equ:cons_alpha}) (\ref{equ:cons_G}) and defined by Algorithm \ref{alg:graph} control FWER at $\alpha$ in the strong sense, then a natural question for drug development is how to obtain the optimal one based on the results from a previous study. Before diving into this optimization problem, we first define an objective function to evaluate different graphs in the following section. 

\subsection{An objective function to evaluate performance}
\label{s:obj}

Remember that in the previous section, we use $\boldsymbol{p}$ to denote the unadjusted $p$-value vector for $m$ endpoints. Given this underlying multivariate data-generating mechanism, we further define an objective function $O \left(\boldsymbol{\alpha}, \boldsymbol{T} \right) $ to measure the performance of a graphical procedure with initial alpha vector $\boldsymbol{\alpha}$ and transition matrix $\boldsymbol{T}$,
\begin{equation}
	\label{equ:obj}
	O \left(\boldsymbol{\alpha}, \boldsymbol{T} \right) = \sum_{i=1}^m v_i \text{ E}_{\boldsymbol{p}} \left\{ D_i \left(\boldsymbol{\alpha}, \boldsymbol{T}, \boldsymbol{p} \right) \right\},
\end{equation}
where the expectation is with respect to the multivariate distribution of $\boldsymbol{p}$, and $v_i$ is pre-specified to represent the relative importance of endpoint $i$ with the constraint $\sum_{i=1}^m v_i = 1$. As a starting point, we focus on the objective function defined in (\ref{equ:obj}) for illustration. In the case study in Section \ref{s:case}, we generalize this objective function to be more clinically meaningful based on the study's objective. We denote the stack of $v_i$'s with the vector $\boldsymbol{v}$. If $v_i = 1/m$ for all $i$'s, then (\ref{equ:obj}) is interpreted as the average of multiplicity adjusted power from all endpoints. In the motivating example, the team can set $\boldsymbol{v} = (0.4, 0.2, 0.3, 0.1)^\prime$ if they treat $H_1$ as the most important target.

Let $\mathcal{A}$ denote the parameter space of $\boldsymbol{\alpha}$, and correspondingly $\mathcal{T}$ for $\boldsymbol{T}$'s. The space $\mathcal{A}$ and $\mathcal{T}$ should satisfy the conditions of a valid graphical approach as in (\ref{equ:cons_alpha}) and (\ref{equ:cons_G}), and be constrained under a specific study design. For example, only $\alpha_1$ and $\alpha_3$ in the motivating example are allowed to be non-zero with sum equal to one-sided FWER $\alpha$. Therefore, $\mathcal{A} = \{(\alpha_1, 0, \alpha_3, 0); \alpha_1 \in [0,\alpha], \alpha_3 \in [0,\alpha], \alpha_1 + \alpha_3 = \alpha \}$. The $\alpha_1$ is a free parameter to be optimized, and further, set $\alpha_3 = \alpha - \alpha_1$. Denote all free parameters in $\boldsymbol{\alpha}$ and $\boldsymbol{T}$ as $\bar{\boldsymbol{\alpha}} \in \bar{\mathcal{A}}$ and $\bar{\boldsymbol{T}} \in \bar{\mathcal{T}}$, respectively. In this motivating example we have $\bar{\mathcal{A}} = \{\alpha_1; \alpha_1 \in [0,\alpha]\}$. There is no inequality constraint on $\alpha_1 \in \bar{\mathcal{A}}$ in this simple problem because $\alpha_3$ is excluded from $\bar{\mathcal{A}}$. In the simulation studies considered in Section \ref{s:sim} and the case study in Section \ref{s:case} with more endpoints, both bounded and inequality constraints exist. 

The goal of our optimization task is to find the optimal graphical approach with $\boldsymbol{\alpha}^{opt}$ and $\boldsymbol{T}^{opt}$ within their corresponding parameter spaces that maximize the object function $O \left(\boldsymbol{\alpha}, \boldsymbol{T} \right)$ in (\ref{equ:obj}),
\begin{align}
	\label{equ:true_obj_opt}
	\left\{\boldsymbol{\alpha}^{opt}, \boldsymbol{T}^{opt} \right\} & =   \argmax_{\boldsymbol{\alpha} \in \mathcal{A}, \boldsymbol{T} \in \mathcal{T}} O \left(\boldsymbol{\alpha}, \boldsymbol{T} \right).
\end{align}
However, $O \left(\boldsymbol{\alpha}, \boldsymbol{T} \right)$ in (\ref{equ:obj}) does not necessarily have a closed form solution due to: (1) the underlying correlation structure in the multivariate distribution of $\boldsymbol{p}$ and (2) the additional dependence in the decision function $D_i \left(\boldsymbol{\alpha}, \boldsymbol{T}, \boldsymbol{p} \right)$ among endpoints introduced by Algorithm \ref{alg:graph}. In practice, a Monte Carlo approach can be implemented to estimate $O \left(\boldsymbol{\alpha}, \boldsymbol{T} \right)$. By simulating $n$ sets of unadjusted $p$-values $\boldsymbol{p}_j = (p_{j1}, p_{j2}, ..., p_{jm}), j = 1, 2, ..., n$, for $m$ endpoints based on prior knowledge, one can use the following working objective function to estimate (\ref{equ:obj}) empirically, 
\begin{equation}
	\label{equ:obj_emp}
	\widehat{O} \left(\boldsymbol{\alpha}, \boldsymbol{T} \right) = \frac{1}{n} \sum_{i=1}^m v_i \sum_{j=1}^n D_i \left(\boldsymbol{\alpha}, \boldsymbol{T}, \boldsymbol{p}_j \right).
\end{equation}
Some standard softwares, for example R package {\it{gMCP}} \citep{gmcp}, can calculate $D_i \left(\boldsymbol{\alpha}, \boldsymbol{T}, \boldsymbol{p}_j \right)$ given each set of simulated unadjusted $p$-values $\boldsymbol{p}_j$. By the Law of Large Numbers, we have
\begin{equation}
	\label{equ:error_obj_central}
	\widehat{O} \left(\boldsymbol{\alpha}, \boldsymbol{T} \right) =  O \left(\boldsymbol{\alpha}, \boldsymbol{T} \right) + o_p(1). 
\end{equation} 
The approximation error of estimating $O \left(\boldsymbol{\alpha}, \boldsymbol{T} \right)$ by $\widehat{O} \left(\boldsymbol{\alpha}, \boldsymbol{T} \right)$ can be arbitrarily small to satisfy practical numerical precision requirements by choosing a sufficiently large $n$ in the Monte Carlo method. In the next section, we introduce our proposed optimization algorithm by using FNN to approximate $\widehat{O} \left(\boldsymbol{\alpha}, \boldsymbol{T} \right)$, and then conduct optimization with available gradient information.

\section{FNN-based optimizer}

In Section \ref{s:fnn}, feedforward neural networks (FNNs) in deep learning are briefly reviewed as powerful representations of complex objective functions. In Section \ref{s:opt_pro}, we illustrate our proposed FNN-based optimization method in detail. It takes advantage of FNN to characterize the non-convex working objective function $\widehat{O} \left(\boldsymbol{\alpha}, \boldsymbol{T} \right)$ and then performs constrained optimization with gradient information. 

\subsection{Feedforward neural networks (FNNs)}
\label{s:fnn}

We first review some basic knowledge of feedforward neural networks (FNNs), which form a very popular and useful set of deep learning models. 

\begin{figure}[h]
	\centering
	\includegraphics[width=0.7\linewidth]{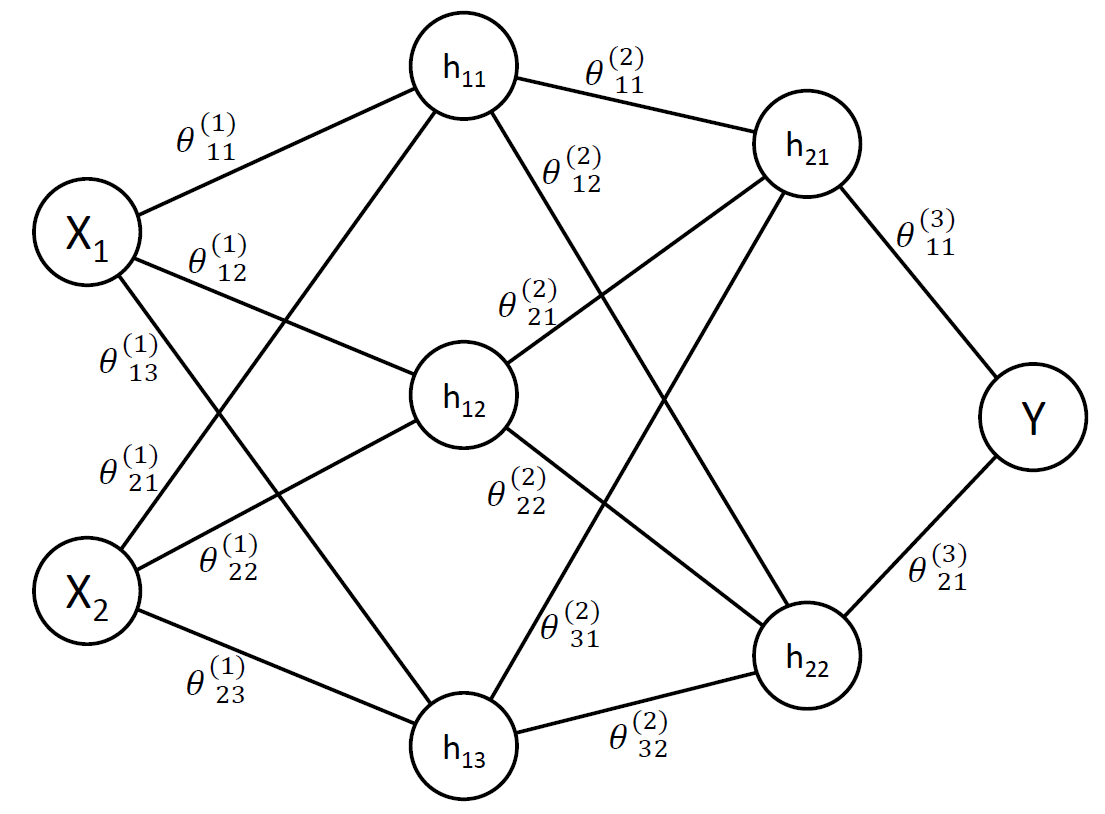}
	\caption{Feedforward neural networks with two hidden layers.}
	\label{F:fnn}
\end{figure}

An FNN defines a mapping $y = f(\boldsymbol{x}; \boldsymbol{\theta})$ and learns the value of parameters $\boldsymbol{\theta}$ that result in the best function approximation with input vector $\boldsymbol{x}$ and output $y$ \citep{goodfellow2016deep}. It typically has four essential components: input data with corresponding targets, layers, loss function and optimizer \citep{deepwithr, liang2018estimating}. Figure $\ref{F:fnn}$ represents an FNN with two hidden layers, which have three and two nodes, respectively. From left to right, input data $\boldsymbol{x}$, which is the vector stack of $x_1$ and $x_2$, are transformed by two hidden layers and further mapped to output target $Y$. The loss function represents the quantity that is minimized during training, for example the cross-entropy for binary classification and mean squared error (MSE) for regression. We choose MSE because our output $\widehat{O} \left(\boldsymbol{\alpha}, \boldsymbol{T} \right)$ ranges from $0$ to $1$. The optimizer determines how the network will be updated based on the loss function. The RMSProp algorithm \citep{hinton2012neural} modifies AdaGrad \citep{duchi2011adaptive} to perform better in the non-convex setting by changing the gradient accumulation into an exponentially weighted moving average. It has been shown to be an effective and practical optimization algorithm for deep neural networks \citep{goodfellow2016deep}, and is used in this article.


For an FNN with $L-1$ hidden layers and one output layer, it can be recursively formulated as 
\begin{equation}
	\label{equ:fnn}
	f(\boldsymbol{x}; \boldsymbol{\theta}) = f^{(L)} \left[ \theta^{(L)\prime} ... f^{(2)} \left\{ \theta^{(2)\prime} f^{(1)}\left(\theta^{(1)\prime} \boldsymbol{x} + \boldsymbol{b}_1 \right) + \boldsymbol{b}_2 \right\} ... +  {b}_L \right]. 
\end{equation}
In the most inner layer, $\theta^{(1)}$ is a weight matrix that transforms input $\boldsymbol{x}$ to the first hidden layer. For example, the dimension of $\theta^{(1)}$ is $2 \times 3$ in Figure \ref{F:fnn}. The number of elements in bias vector $\boldsymbol{b}_1$ is equal to the number of nodes in the first layer (i.e., $3$). The vector $\boldsymbol{\theta}$ denotes a stack of all those weight and bias parameters. There are many choices for the activation function $f^{(1)}()$, for example, the rectified linear unit or ReLU \citep{nair2010rectified}, the softplus function \citep{dugas2001incorporating}, and the sigmoid function. 

The approximation error of an FNN $f(\boldsymbol{x}; \boldsymbol{\theta})$ in approximating the objective function $O(\boldsymbol{x})$ in (\ref{equ:obj}) is defined as 
\begin{equation*}
	\sup_{\boldsymbol{x}} \Big|f(\boldsymbol{x}; \boldsymbol{\theta}) - O(\boldsymbol{x}) \Big|,
\end{equation*}
where $\boldsymbol{x} = (\boldsymbol{\alpha}, \boldsymbol{T})$. Previously, it has been shown that a depth-2 neural network with sigmoid activation function can approximate any continuous function to a desired accuracy, with sufficiently large number of nodes \citep{cybenko1989approximation}. Since then, interest has shifted towards a deeper network, since the multilayer feedforward architecture itself gives neural networks the potential of being universal approximators \citep{hornik1991approximation}. Recently, \cite{bach2017breaking} provides the uniform approximation error of Lipschitz-continuous functions in the context of high-dimensional non-linear variable selection. The error bound for approximation functions in Sobolev spaces by deep ReLU networks is studied by \cite{yarotsky2017error}. 

As discussed in Section \ref{s:obj}, theoretical properties of the objective function $O \left(\boldsymbol{\alpha}, \boldsymbol{T} \right)$ in (\ref{equ:obj}) is hard to study, mainly due to the dependence between decision function $D_i \left(\boldsymbol{\alpha}, \boldsymbol{T}, \boldsymbol{p} \right)$'s introduced by Algorithm \ref{alg:graph}. In this article, we use simulation studies to empirically check the MSE of modeling the working objection function $\widehat{O}(\boldsymbol{\alpha}, \boldsymbol{T})$ by FNN, with more details in the following Section \ref{s:train}. Moreover, our algorithm has a last step in Section \ref{s:fine_tune} to fine tune the optimal solution obtained from the FNN model to correct approximation errors.





\subsection{Optimizing procedures}
\label{s:opt_pro}

In this section, we illustrate our optimizing procedures in six steps.

\subsubsection{Define an objective function}
\label{s:define}
The first step is to specify an objective function $O \left(\boldsymbol{\alpha}, \boldsymbol{T} \right)$ to measure the performance of the graphical procedure for MTP. The vector $\boldsymbol{v}$ in (\ref{equ:obj}) needs to be pre-specified and reflects the relative importance of different endpoints.  

\subsubsection{Obtain training data}
\label{s:data}
The second step is to generate training data with $B$ graphs and their corresponding objective functions (\ref{equ:obj}). In each graph $b$, one randomly generates $\boldsymbol{\alpha}_b \in \mathcal{A}$ and $\boldsymbol{T}_b \in \mathcal{T}$ under conditions (\ref{equ:cons_alpha}) and (\ref{equ:cons_G}), along with other constraints based on different study objectives. In the motivating example we have $\mathcal{A} = \{(\alpha_1, 0, \alpha_3, 0), \alpha_1 \in [0,\alpha], \alpha_3 \in [0,\alpha], \alpha_1 + \alpha_3 = \alpha \}$. In this case, $\alpha_1$ can be sampled from a uniform distribution $Unif(0, \alpha)$ and further set $\alpha_3 = \alpha - \alpha_1$. The free parameter vector $\bar{\boldsymbol{\alpha}}_b$ only contains $\alpha_1$ in this case. It is important to enforce these constraints at this stage to achieve constraint optimization of the graphical approach.

We further simulate $n$ sets of unadjusted $p$-values $\boldsymbol{p}_i = (p_{i1}, p_{i2}, ..., p_{im}), i = 1, 2, ..., n$ based on prior knowledge. Suppose that the marginal powers of four endpoints are $95\%$, $88\%$, $92\%$ and $85\%$, which correspond to a test statistic's mean at $\boldsymbol{e} = (3.60, 3.13, 3.37, 3.00)$ with one-sided type I error $\alpha = 0.025$. We adopt a popular assumption that the test statistics from $m$ hypotheses follow a multivariate normal distribution \citep{dmitrienko2013traditional, bretz2016multiple}. Unit variance is assumed for demonstration. Without loss of generality, by assuming that a larger statistic corresponds to a better clinical outcome, the one-sided $p$-value is calculated as the upper cumulative distribution function from a standard normal distribution. Having $\boldsymbol{p}_i$'s simulated, one calculates $\widehat{O} \left(\boldsymbol{\alpha}_b, \boldsymbol{T}_b \right)$ in (\ref{equ:obj_emp}) for each graph $b$. The input covariate vector $\bar{\boldsymbol{x}}_b$ of FNN is $(\bar{\boldsymbol{\alpha}}_b, \bar{\boldsymbol{T}}_b)$, while the output variable is $\widehat{O} \left(\boldsymbol{\alpha}_b, \boldsymbol{T}_b \right)$. The dimension of $\bar{\boldsymbol{x}}_b$ is equal to the number of input parameters of FNN on the left hand side of Figure \ref{F:fnn}. 

\subsubsection{Select FNN structure}
\label{s:cross}

The next step is to select the structure of FNN, specifically the width (number of nodes), the depth (number of layers), and the rate of the dropout technique, which randomly deactivates a certain proportion of nodes in each iteration, to accommodate the potential overfitting issue in FNN. 

The most common practice is to perform a $k$-fold cross-validation procedure on several reasonable candidate structures \citep{goodfellow2016deep}. In cross-validation, a partition of the dataset is formed by splitting it into $k$ non-overlapping subsets. On each trial $i$, for $i = 1, ..., k$, the $i$-th subset of data is used as the validation set while the rest of the data is used as the training set. The validation error is calculated by averaging test error across $k$ trials. We let $k = 5$ to implement a $5$-fold cross validation. The final FNN structure is selected as the one with the smallest training error among candidates. Validation error on other measures can also be utilized, and the performance of our method is consistent.

We recommend starting with an architecture with a relatively large capability to reduce the training error (MSE) under a desired level of tolerance, for example $10^{-4}$. This ensures that the functional space defined by the structure is large enough to include the underlying objection function, or a very good approximation of it. However, this structure may be overwhelmed with a high validation error. Then we apply the dropout technique as a regulation approach to prevent overfitting and to increase the generalizability of the model. In the context of this article, exploratory simulations show that the performance of our FNN-based optimizer is robust to different choices of FNN structures, when both the training MSE and the validation MSE are less than $10^{-4}$.   

\subsubsection{Train FNN}
\label{s:train}
The following step is to train the FNN with structure obtained in 5-fold cross validation with input covariates $\bar{\boldsymbol{x}}_b$ and output $\widehat{O}(\boldsymbol{\alpha}_b, \boldsymbol{T}_b)$, $b = 1, 2, ..., B$. Covariates $\bar{\boldsymbol{x}}_b$ are standardized to achieve better performance of a gradient-based optimizer, and are further transformed back to the original scale after fitting. Mean squared error (MSE) is utilized as the loss function,
\begin{equation}
	\label{equ_mse_FNN}
	\frac{1}{B} \sum_{b=1}^B \left[\widehat{O}(\boldsymbol{\alpha}_b, \boldsymbol{T}_b) - f(\bar{\boldsymbol{x}}_b; {\boldsymbol{\theta}}) \right]^2.
\end{equation}
The least squares estimator $\boldsymbol{\widehat{\theta}}$ is obtained from the RMSProp Algorithm discussed in Section \ref{s:fnn} as the $\boldsymbol{\theta}$ that minimizes this loss function. The fitted FNN is denoted as $f(\bar{\boldsymbol{x}}; \widehat{\boldsymbol{\theta}})$. For a specific problem, it is critical to check this MSE to evaluate the approximation error of FNN empirically. The estimation error between $\widehat{O}(\boldsymbol{\alpha}_b, \boldsymbol{T}_b)$ and ${O}(\boldsymbol{\alpha}_b, \boldsymbol{T}_b)$ is further controlled by using a relatively larger $n$ in (\ref{equ:error_obj_central}). Before implementing our proposed optimization method, it is critical to quantify the approximation ability of FNN by checking the MSE in (\ref{equ_mse_FNN}).  

Since sigmoid functions saturate (have small gradients) when input data are at two tails, we further normalize $\widehat{O}(\boldsymbol{\alpha}_b, \boldsymbol{T}_b)$ to a subset of $[0, 1]$, for example $[0.3, 0.7]$. The optimal solution would be invariant under this transformation. The whole training process is implemented by the R interface {\it{keras}} \citep{rkeras, rtensorflow} to a high-level neural networks API {\it{Keras}} \citep{chollet2015keras} with back-end engine {\it{Tensorflow}} \citep{tensorflow2015whitepaper} developed by Google Inc. We set the training epoch as $10^3$. 

\subsubsection{Perform constrained optimization}
\label{s:perform}
Up to this point, we have transformed the original optimization problem in (\ref{equ:true_obj_opt}) to the following constrained minimization problem of identifying optimal solution $\bar{\boldsymbol{x}}^{opt}_b$ on  $-f(\bar{\boldsymbol{x}}_b; \widehat{\boldsymbol{\theta}})$, 
\begin{align}
	\bar{\boldsymbol{x}}^{opt}_b & =   \argmax_{\bar{\boldsymbol{\alpha}} \in \bar{\mathcal{A}}, \bar{\boldsymbol{T}} \in \bar{\mathcal{T}}} f(\bar{\boldsymbol{x}}_b; \widehat{\boldsymbol{\theta}})  =  \argmin_{\bar{\boldsymbol{\alpha}} \in \bar{\mathcal{A}}, \bar{\boldsymbol{T}} \in \bar{\mathcal{T}}} \left\{ -f(\bar{\boldsymbol{x}}_b; \widehat{\boldsymbol{\theta}}) \right\}. \label{equ:opt_graph}
\end{align}
The optimal graph parameters $\boldsymbol{\alpha}^{opt}$ and $\boldsymbol{T}^{opt}$ are further calculated from $\bar{\boldsymbol{x}}^{opt}_b$. 

Since $-f(\bar{\boldsymbol{x}}_b; \widehat{\boldsymbol{\theta}})$ in (\ref{equ:fnn}) is not necessarily a convex function, then the Karush-Kuhn-Tucker (KKT)
conditions \citep{karush1939minima,kuhn1951nonlinear} are not sufficient for a point to be globally optimal. Even with gradient information available, finding the global optimal solution is still challenging depending on the objective function at hand \citep{kramer2011derivative}. We turn to the augmented Lagrangian method \citep{hestenes1969multiplier,powell1969method}, which seeks the solution by replacing the original constrained problem by a sequence of unconstrained subproblems \citep{nocedal2006nonlinear}. This algorithm is related to the quadratic penalty method \citep{courant1943variational}, but reduces the possibility of ill conditioning of the subproblems by introducing a Lagrange multiplier into the function to be minimized. 

This algorithm, as well as COBYLA and ISRES discussed later on, are implemented by the R package {\it{nloptr}} \citep{nloptr}, which is the R interface to a nonlinear optimization library {\it{NLopt}} \citep{nlopt,conn1991globally,birgin2008improving}. The fractional tolerance on the input data is $10^{-5}$, which means that the algorithm terminates when the changes of each parameter in one iteration are less than $10^{-5}$ multiplied by the absolute value of the parameter. The maximum number of iterations is $10^5$. 

\subsubsection{Fine tune the final optimal solution}
\label{s:fine_tune}

As a final step, we fine-tune the solution with COBYLA, an existing derivative-free optimization method that can handle inequality constraints. Essentially, our optimal solution from the previous step is used as the starting values in COBYLA. The fractional tolerance on the input data is $10^{-4}$, and the maximum number of iterations is $10^4$. 



\section{Simulation studies}
\label{s:sim}

Now we move on to a simulation study to evaluate the performance of our proposed FNN-based optimizer against the stochastic search method (SSM) and two derivative-free optimization methods that can handle bound and inequality constraints: COBYLA and ISRES.

Suppose that the study objective is to identify the optimal graphical procedure that maximizes a weighted average of multiplicity adjusted power for $m=6$ endpoints. One can work out that the input covariate vector $\bar{\boldsymbol{x}}$ of FNN is $(\alpha_1, \alpha_2, \alpha_3, \alpha_4, \alpha_5, T_{12}, T_{13}, T_{14}, T_{15}$, $ T_{21}, T_{23}, T_{24}, T_{25}, T_{31}, T_{32}$, $T_{34}, T_{35}, T_{41}, T_{42}, T_{43}, T_{45}, T_{51}, T_{52}, $ $T_{53}, T_{54}, T_{61}, T_{62}, T_{63}, T_{64})$ with $29$ elements. The constraints are:
\begin{align}
	& 0 \leq \alpha_j \leq \alpha  \text{ for } j \in \{1, 2, 3, 4, 5\}, \label{sim_cons_1} \\
	& \sum_{j \in \{1, 2, 3, 4, 5\}} \alpha_j \leq \alpha, \label{sim_cons_2} \\
	& 0 \leq T_{ij} \leq 1  \text{ for } i \in \{1, 2, 3, 4, 5\}, j \in \{1, 2, 3, 4, 5, j \neq i\},
	\:\:\:\:\:\: 0 \leq T_{6j} \leq 1, j \in \{1, 2, 3, 4\}, \label{sim_cons_3} \\
	& \sum_{j \in \{1, 2, 3, 4, 5, j \neq i\}} T_{ij} \leq 1  \text{ for } i \in \{1, 2, 3, 4, 5\} \:\:\:\:\:\:
	\sum_{j \in \{1, 2, 3, 4\}} T_{6j} \leq 1.
	\label{sim_cons_4}
\end{align}
Condition (\ref{sim_cons_1}) says that the initial significance level from each of the first $5$ endpoints is bounded between $0$ and FWER at $\alpha$, while constraint (\ref{sim_cons_2}) ensures this for the last endpoint because $\alpha_6 = \alpha - \sum_{j=1}^5 \alpha_j$. Constraints (\ref{sim_cons_3}) and (\ref{sim_cons_4}) are the corresponding constraints for each of the $6$ rows in the transitional matrix $\boldsymbol{T}$.  

We consider $\boldsymbol{v} = (0.3, 0.3, 0.1, 0.1, 0.1, 0.1)$ as the relative importance in (\ref{equ:obj}) in this section, and turn to a different $\boldsymbol{v}$ in the Section \ref{s:case} case study. As discussed in Section \ref{s:data}, we assume that the test statistics from $m$ endpoints follow a multivariate normal distribution with unit variance and mean computed from their corresponding marginal power under a one-sided FWER at $0.025$. The setup parameters are specified in Table \ref{t:sim_sens_parameters} with varying marginal power, different correlation structures and varying magnitudes of correlation. 

\begin{table}[ht]
	\centering
	\footnotesize
	\caption{Parameter specifications for simulations}
	\label{t:sim_sens_parameters}
	\begin{tabular}{cccc}
		\midrule
		Scenario & Marginal power & Correlation structure & Correlation magnitude   \\ 
		\midrule
		$L_1$ & $(0.8, 0.8, 0.6, 0.6, 0.4, 0.4)$ & Compound symmetry & $0$ \\
		$L_2$ &  &  & $0.3$ \\
		$L_3$ &  &  & $0.5$ \\
		\\
		$L_4$ & $(0.9, 0.9, 0.8, 0.8, 0.6, 0.6)$ & Compound symmetry & $0.3$ \\
		$L_5$ &  & AR(1) &  \\
		$L_6$ &  & Banded Toeplitz &  \\
		\\
		$L_7$ & $(0.9, 0.8, 0.7, 0.6, 0.5, 0.4)$ & Compound symmetry & $0.3$ \\
		$L_8$ & $(0.9, 0.9, 0.7, 0.7, 0.6, 0.6)$ &  &  \\
		$L_9$ & $(0.95, 0.95, 0.8, 0.8, 0.6, 0.6)$ &  &  \\
		\midrule
	\end{tabular}
\end{table} 

For the FNN-based optimizer as described in Section \ref{s:opt_pro}, we simulate $B = 10^3$ random graphs and $n = 10^6$ sets of $p$-values to establish the training dataset. The size $B=10^3$ is sufficient to give us the training MSE and the validation MSE less than $10^{-4}$ in all scenarios considered, but it can be increased in more complicated cases. In cross-validation while selecting the FNN structure, the following $6$ sets of candidate structures are considered: $2$ layers with drop-out rate $0$, $3$ layers with rate at $0$, $4$ layers with rate $0$, $2$ layers with rate $0.3$, $3$ layers with rate $0.3$ and $4$ layers with rate $0.3$. The number of nodes per layer is considered at $30$. This cohort of FNN candidate skeletons are utilized throughout this article. 

In ISRES and COBYLA, fractional tolerance on the input data is $10^{-4}$, which is consistent with the termination condition at our fine-tuning step at Section \ref{s:fine_tune}. The maximum evaluation time is set as $1.5$ times the fitting time of the FNN-based optimizer as described in Section \ref{s:opt_pro}. The initial values are randomly generated under the constraints in (\ref{sim_cons_1}), (\ref{sim_cons_2}), (\ref{sim_cons_3}) and (\ref{sim_cons_4}). We consider a size of $10^3$ for the random search in the stochastic search method (SSM). This is equal to the training size $B$ in our method. 

In Table \ref{t:sim}, we summarize the optimal working objective function $\widehat{O} \left(\boldsymbol{\alpha}, \boldsymbol{T} \right)$ identified by the FNN-based optimizer, ISRES, COBYLA and SSM along with their corresponding convergence times in minutes based on a MacBook Pro with $2.3$ GHz Intel Core i7. The average from $5$ separate optimizations are reported for FNN-based method, ISRES and COBYLA to evaluate the robustness of their performance. Our method has a smaller standard deviation at $0.14\%$, compared with COBYLA at $0.72\%$ and ISRES at $1.09\%$, averaging the standard deviations across $9$ scenarios. In all scenarios evaluated, our FNN method consistently identifies a graph with the highest $\widehat{O} \left(\boldsymbol{\alpha}, \boldsymbol{T} \right)$. However, the performance of the other two methods is not stable; for example COBYLA finds $63.7\%$ compared to $64.2\%$ from the FNN method in $L_7$, and ISRES yields $48.2\%$ compared to $59.1\%$ in $L_3$. The SSM method can also lead to substantial optimal power loss in some scenarios. For example, the deviance can be as high as $5.1\%$ compared to our FNN method in $L_3$ ($59.1\%$ versus $54.0\%$). As for convergence, SSM is the fastest, and COBYLA is the second fastest, followed by our FNN-based method, followed by ISRES. The convergence time of ISRES is missing because it does not converge before the maximum wall time, which is the computational time of the FNN-based optimizer multiplied by $1.5$. Our FNN-based optimizer offers a better balance between time efficiency and robustness in identifying the optimal graphical approach. 

\begin{table}[ht]
	\centering
	\footnotesize
	\caption{Optimal $\widehat{O} \left(\boldsymbol{\alpha}, \boldsymbol{T} \right)$ identified by FNN, ISRES, COBYLA and SSM with the maximum solution highlighted}
	\label{t:sim}
	\begin{tabular}{ccccccccc}
		\midrule
		Scenario & \multicolumn{4}{c}{Optimal $\widehat{O} \left(\boldsymbol{\alpha}, \boldsymbol{T} \right)$ }& \multicolumn{4}{c}{Convergence time (minutes)}\\ 
		\cmidrule(lr){2-5}\cmidrule(lr){6-9}
		& FNN & COBYLA & ISRES  & SSM & FNN & COBYLA & ISRES & SSM \\ 
		\midrule
		$L_1$ & {\underline{\bf 55.9\%}} & 55.5\% & 47.8\% & 53.4\% & 25.0 & 13.8 & - & 3.6 \\ 
		$L_2$ & {\underline{\bf 57.9\%}} & 57.4\% & 48.7\% & 55.6\% & 29.2 & 18.2 & - & 4.5 \\ 
		$L_3$ & {\underline{\bf 59.1\%}} & 58.8\% & 48.2\% & 54.0\% & 30.8 & 20.2 & - & 3.8 \\ 
		\\
		$L_4$ & {\underline{\bf 74.6\%}} & 74.4\% & 68.9\% & 72.6\% & 30.2 & 18.3 & - & 4.4 \\ 
		$L_5$ & {\underline{\bf 74.0\%}} & 73.8\% & 69.7\% & 72.7\% & 28.0 & 17.4 & - & 4.5 \\ 
		$L_6$ & {\underline{\bf 74.0\%}} & 73.7\% & 69.7\% & 72.9\% & 28.0 & 15.2 & - & 4.6 \\
		\\
		$L_7$ & {\underline{\bf 64.2\%}} & 63.7\% & 55.6\% & 61.5\% & 30.7 & 18.7 & - & 4.1 \\ 
		$L_8$ & {\underline{\bf 71.8\%}} & 71.6\% & 65.7\% & 69.3\% & 30.6 & 20.3 & - & 4.4 \\ 
		$L_9$ & {\underline{\bf 79.4\%}} & 79.2\% & 74.6\% & 78.0\% & 34.3 & 23.8 & - & 4.7 \\ 
		\midrule
	\end{tabular}
\end{table}

We observe that ISRES does not converge within the given wall time, and further delivers a low optimal value. A possible reason is that it takes longer for ISRES to comprehensively walk through the whole parameter space in this setup with a moderate dimension of $\bar{\boldsymbol{x}}$ at $29$. The performance of COBYLA is not stable in those settings, as demonstrated by the relatively larger standard deviation and smaller mean optimal values compared with FNN. The reason can be that COBYLA is more likely to get stuck in the local optimal. Our proposed method, on the other hand, first seeks a parametric surrogate function approximating the working objective function in (\ref{equ:obj_emp}) by FNN, and then performs optimization with available gradient information. Therefore, our FNN-based optimizer consistently achieves the highest power across all scenarios. 

From a practical point of view, both FNN and COBYLA have satisfactory performance based on this simulation study. COBYLA takes a moderate computational time of approximately $20$ minutes to deliver the solution. Our proposed FNN-based method offers an alternative option to further enhance the power. An extra $15$ minutes is tolerable as compared with study duration over years in confirmatory trials. Moreover, even a fraction of a percent of power gain is non-trivial considering the high cost of clinical trials. More discussions on this are provided in Section \ref{sec:conclu}. 

\begin{figure}[h]
	\centering
	\includegraphics[width=0.9\linewidth]{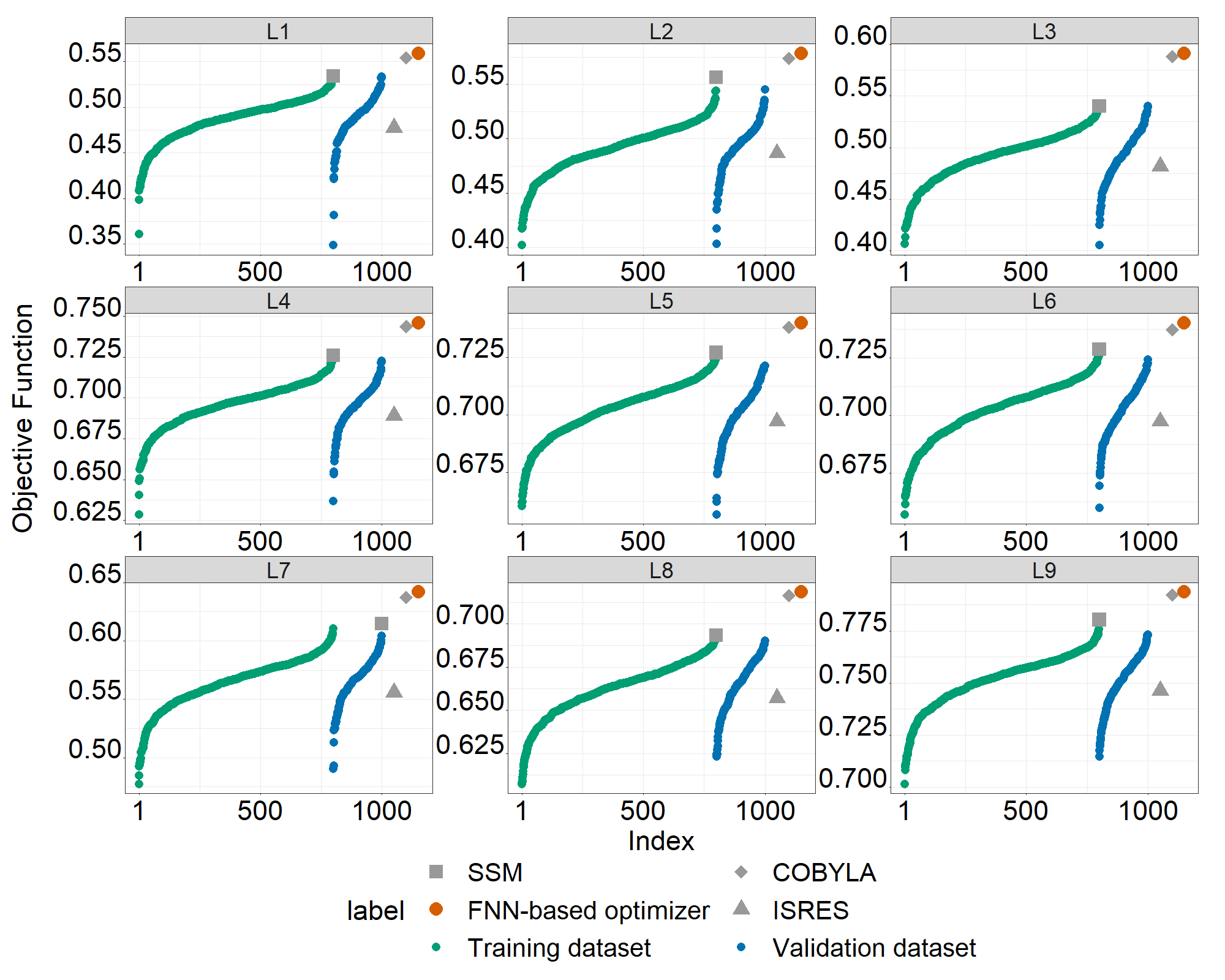}
	\caption{The optimal objective function identified by FNN, COBYLA, ISRES and SSM}
	\label{F:sim_plot}
\end{figure}

In Figure \ref{F:sim_plot}, we visualize the performance of our FNN-based method and the other three comparators. In each scenario, we plot $800$ training datasets in green in the order of their working objective functions from small to large, and then $200$ validation datasets with blue in order. The maximum of both training and validation datasets is the solution of SSM. The optimal graph identified by ISRES in triangle, COBYLA in rhombus and FNN in orange circle are plotted on the right. Next, we evaluate the residuals of utilizing FNN to approximate $\widehat{O}(\boldsymbol{\alpha}, \boldsymbol{T})$, which is  $\widehat{O}(\boldsymbol{\alpha}, \boldsymbol{T}) - f(\bar{\boldsymbol{x}}; \widehat{\boldsymbol{\theta}})$ as in (\ref{equ_mse_FNN}). The residuals from the left $800$ training datasets are generally smaller than those from the right $200$ validation datasets. 

\begin{figure}[h]
	\centering
	\includegraphics[width=0.9\linewidth]{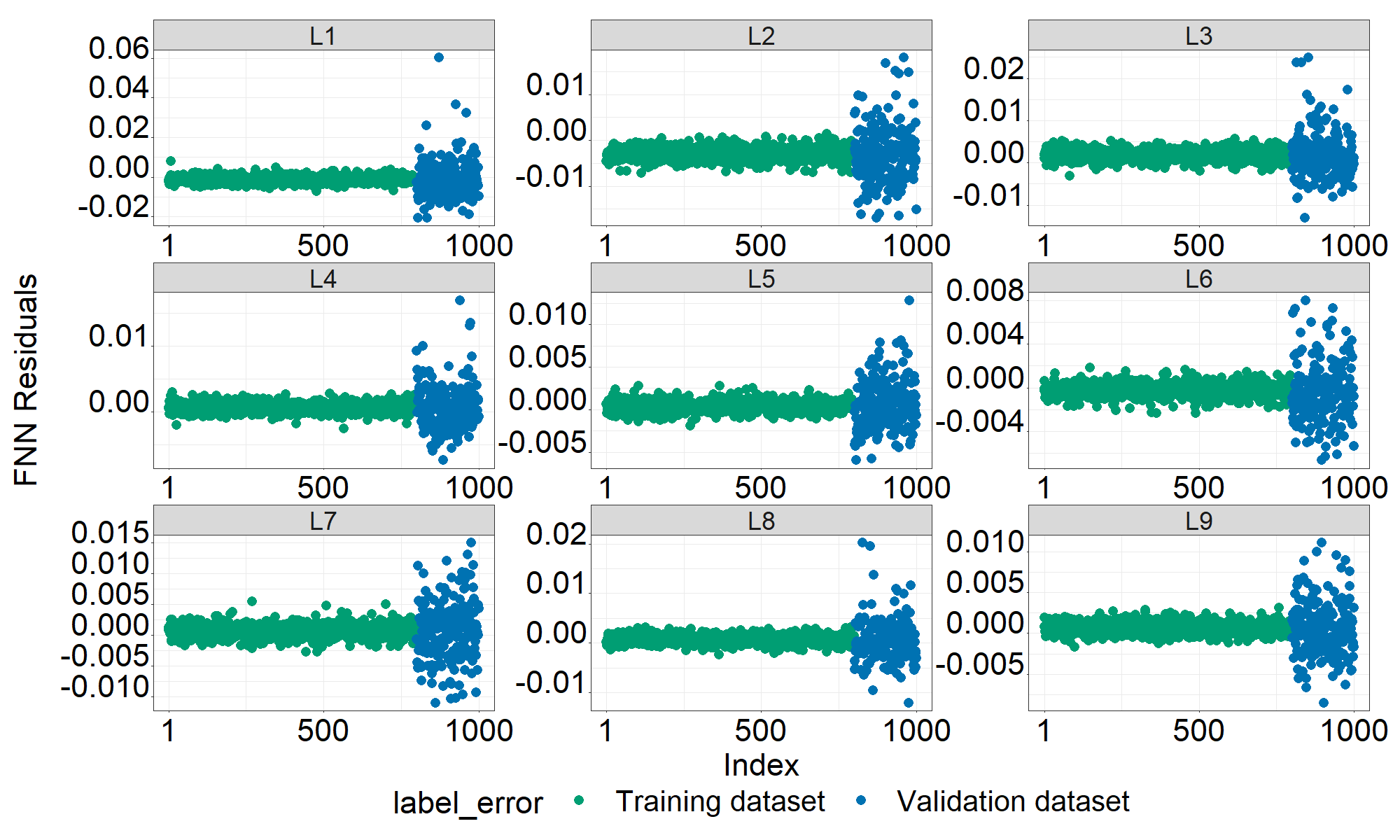}
	\caption{FNN residuals of estimating the working objective function}
	\label{F:sim_error_plot}
\end{figure}

\section{A case study}
\label{s:case}

In this section, we apply our FNN-based optimizing approach to a generic study with one primary endpoint, denoted as $H_1$, and four secondary endpoints, denoted as $H_2$, $H_3$, $H_4$ and $H_5$. This example is particularly relevant because while testing the primary endpoint first is clear, the tactic for testing the secondary endpoints can be flexible. Since the primary endpoint is tested first, we fix the first element in $\boldsymbol{\alpha}$ at the one-sided FWER $0.025$ and the remaining components at $0$. There is no element in $\bar{\boldsymbol{\alpha}}_b$ to be optimized. In the transition matrix $\boldsymbol{T}$, $H_1$ can freely pass its error rate to all secondary endpoints, and each secondary endpoint can recycle theirs to the other $3$ secondary endpoints but not back to the primary endpoint since it must have been rejected first. Therefore, there are $3 + 4 \times 2 = 11$ elements in $\bar{\boldsymbol{T}}_b = (T_{12}, T_{13},T_{14},T_{23},T_{24},T_{32},T_{34},T_{42},T_{43},T_{52},T_{53})$, and each element is bounded between $0$ and $1$. The additional constraints are:
\begin{align*}
	T_{12} + T_{13} + T_{14} & \leq 1,\\
	T_{23} + T_{24}  & \leq 1, \\
	T_{32} + T_{34}  & \leq 1, \\
	T_{42} + T_{43}  & \leq 1, \\
	T_{52} + T_{53}  & \leq 1.
\end{align*}

Suppose that the study team assigns $v_2 = 0.6$, $v_3 = 0.2$, $v_4 = v_5 = 0.1$ to the following objective function,
\begin{equation}
	\label{equ:case1_obj}
	O^\prime \left(\boldsymbol{\alpha}, \boldsymbol{T} \right) = \sum_{i=2}^{5} v_i
	\text{ E}_{\boldsymbol{p}} \left\{ D^\prime_i \left(\boldsymbol{\alpha}, \boldsymbol{T}, \boldsymbol{p} \right) \right\}, 
\end{equation}
where $ D^\prime_i \left(\boldsymbol{\alpha}, \boldsymbol{T}, \boldsymbol{p} \right)=1$ if both $H_i$ and $H_1$ are rejected by the graphical approach $g(\boldsymbol{\alpha}, \boldsymbol{T})$, and $0$ otherwise, for $i = 2, 3, 4, 5$. This reflects the clinical interpretation that the rejection of a secondary endpoint is only meaningful if the primary endpoint has been rejected. Note that we exclude the adjusted power of the primary endpoint in equation (\ref{equ:case1_obj}), because the optimizer is equivalent given the constraints on $\boldsymbol{\alpha}$ in the study setup. 

We further assume that the test statistics follow a multivariate normal distribution with a compound symmetric structure and a common correlation at $0.5$. The marginal power of the primary endpoint is assumed to be $95\%$, and $90\%$, $85\%$, $65\%$ and $60\%$ for secondary endpoints. The parameters in the FNN-based optimizer, ISRES and COBYLA are the same as those specified in Section \ref{s:opt_pro} and \ref{s:sim}.

\begin{figure}[h]
	\centering
	\includegraphics[width=0.9\linewidth]{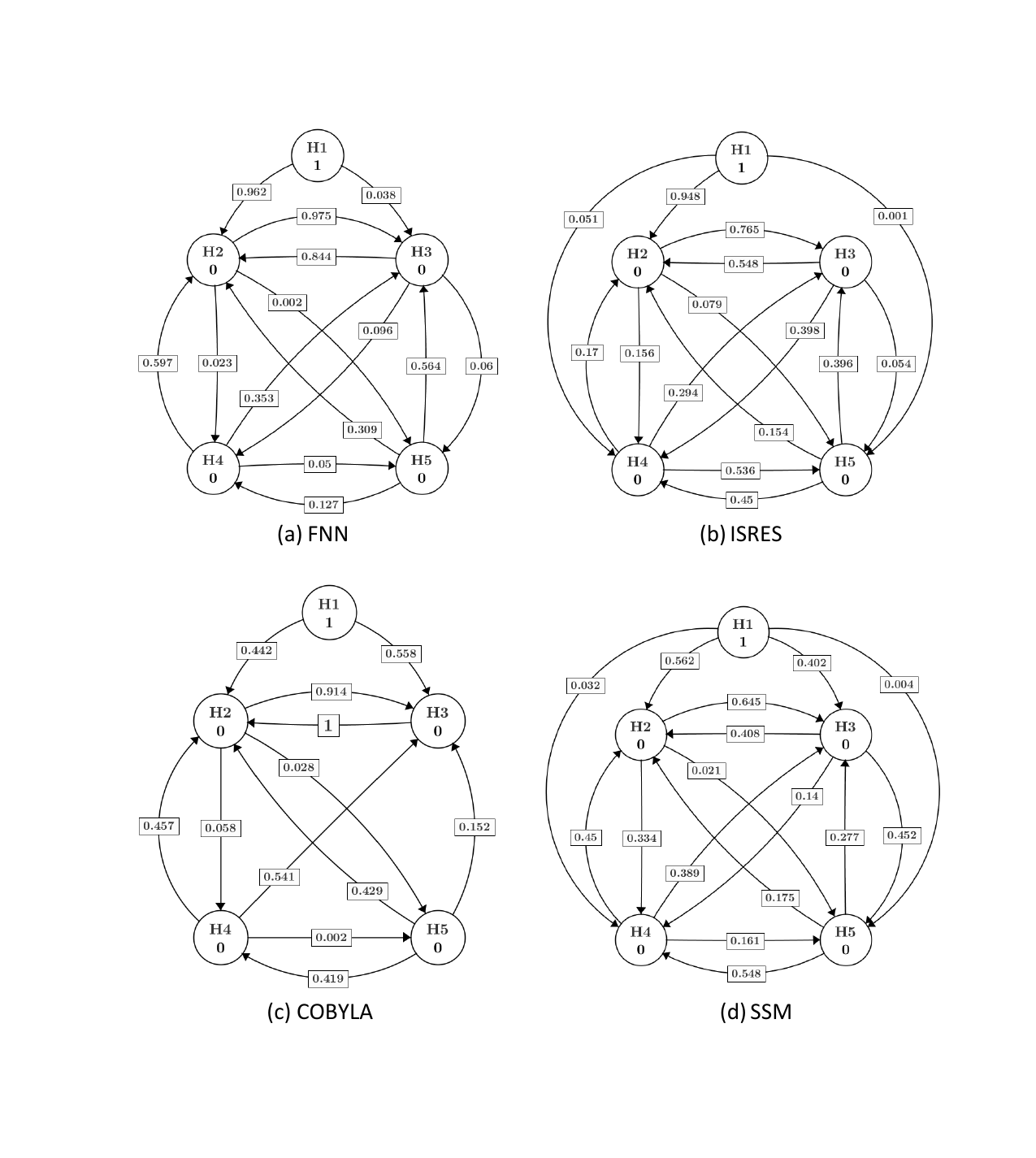}
	\caption{Optimal graph identified by FNN, ISRES, COBYLA and SSM}
	\label{F:case_visu}
\end{figure}

In Table \ref{t:case}, we list the optimal $\widehat{O}^\prime \left(\boldsymbol{\alpha}, \boldsymbol{T} \right)$ and the multiplicity adjusted power of each endpoint from the FNN-based method, ISRES, COBYLA and SSM. Our method achieves the highest objective function at $78.0\%$, which is approximately $0.6\%$ to $1.4\%$ higher than the other three methods. When it comes to convergence time, COBYLA takes $4.0$ minutes, which is shorter than the $17.1$ minutes from our method. ISRES does not converge in the given wall time at $25.7$ minutes. The optimal graphs identified by the four methods are also visualized in Figure \ref{F:case_visu}. To demonstrate the reproducibility of our findings, we further perform $100$ replications of this case study. Our method has the highest mean of optimal $\widehat{O}^\prime (\boldsymbol{\alpha}, \boldsymbol{T})$ at $78.0\%$, compared with $77.0\%$ from ISRES, $77.7\%$ from COBYLA, and $76.4\%$ from SSM. The results are consistent with our report at Table \ref{t:case}. Our proposed method also has a relatively small standard deviation at $0.04\%$, while ISRES has $0.23\%$, COBYLA has $0.54\%$, and SSM has $0.03\%$. 

\begin{table}[ht]
	\centering
	\scriptsize
	\caption{Optimal $\widehat{O}^\prime \left(\boldsymbol{\alpha}, \boldsymbol{T} \right)$ and $\widehat{\text{E}}_{\boldsymbol{p}} \left\{ D^\prime_i \left(\boldsymbol{\alpha}, \boldsymbol{T}, \boldsymbol{p} \right) \right\}$ identified by FNN, ISRES, COBYLA and SSM}
	\label{t:case}
	\begin{tabular}{ccccccc}
		\midrule
		Method & $\widehat{O}^\prime \left(\boldsymbol{\alpha}, \boldsymbol{T} \right)$ & $\widehat{\text{E}}_{\boldsymbol{p}} \left\{ D^\prime_1 \left(\boldsymbol{\alpha}, \boldsymbol{T}, \boldsymbol{p} \right) \right\}$ & $\widehat{\text{E}}_{\boldsymbol{p}} \left\{ D^\prime_2 \left(\boldsymbol{\alpha}, \boldsymbol{T}, \boldsymbol{p} \right) \right\}$ & $\widehat{\text{E}}_{\boldsymbol{p}} \left\{ D^\prime_3 \left(\boldsymbol{\alpha}, \boldsymbol{T}, \boldsymbol{p} \right) \right\}$ & $\widehat{\text{E}}_{\boldsymbol{p}} \left\{ D^\prime_4 \left(\boldsymbol{\alpha}, \boldsymbol{T}, \boldsymbol{p} \right) \right\}$ & $\widehat{\text{E}}_{\boldsymbol{p}} \left\{ D^\prime_5 \left(\boldsymbol{\alpha}, \boldsymbol{T}, \boldsymbol{p} \right) \right\}$ \\ 
		\midrule
		FNN & 78.0\% & 95.0\% & 86.7\% & 78.4\% & 54.6\% & 48.5\% \\ 
		ISRES & 77.4\% & 95.0\% & 86.6\% & 75.2\% & 56.8\% & 47.6\% \\ 
		COBYLA & 77.2\% & 95.0\% & 84.7\% & 80.8\% & 54.5\% & 48.4\% \\ 
		SSM & 76.6\% & 95.0\% & 84.0\% & 79.0\% & 54.4\% & 49.7\% \\ 
		\midrule
	\end{tabular}
\end{table}

%
%

\section{Concluding remarks}
\label{sec:conclu}

In this manuscript, we propose an FNN-based optimization framework for the graphical procedure of multiplicity control in confirmatory clinical trials. This framework takes advantage of the strong functional representation of deep neural networks and further utilizes constraint optimization techniques to locate the solution. Simulation studies show that our FNN-based optimizer consistently identifies the optimal graph, and has a better balance between robustness and time efficiency as compared to two popular derivative-free optimization methods that can handle bound and inequality constraints. 

Our proposed method numerically approximates the optimal graph from the graphical approach with respect to the objective function under specified constraints. An optimal solution may not be unique. Numerical approximation is deemed appropriate due to the intractable nature of the underlying objective function. Numerical precision needs to be considered case by case because of the different number of simulated graphs ($B$) and finite simulated $p$-values ($n$) in the training dataset. For increased precision of the approximate solution, one can further increase $B$ and $n$.

In practice, a relatively simplified graph may be more palatable for the clinical team, as compared to the numerically optimized upper bound with respect to the objective function identified by our method. If the distance in the objective function is relatively small, then this evidence adds more justification to the usage of the proposed simple graph. On the other hand, if power is of main interest, then our method has moderate power gain as compared with the two existing derivative-free methods and the stochastic search method (SSM) demonstrated by the simulation studies and the case study. As shown in Table \ref{t:sim}, the average multiplicity adjusted power increase can be as high as $5.1\%$ compared with the SSM method, over $10\%$ as compared with ISRES, and $0.5\%$ compared with COBYLA. This makes our proposed method appealing in confirmatory studies where several secondary endpoints are targeted for labeling purposes. Even though the gain in some cases is merely a fraction of a percent of power, it is still worth the additional computing time, which is never more than a couple of minutes, especially if either the cost of the study is high or the stakes are high based on participation of subjects with serious afflictions.

\section*{Acknowledgment}

The support of this manuscript was provided by AbbVie Inc. and Takeda Pharmaceuticals USA, Inc. AbbVie Inc. and Takeda Pharmaceuticals USA, Inc. participated in the review and approval of the content. Tianyu Zhan is an employee of AbbVie Inc. Alan Hartford is employed by Takeda Pharmaceuticals USA, Inc. Jian Kang is Professor in the Department of Biostatistics at the University of Michigan, Ann Arbor. Kang's research was partially supported by NIH R01 GM124061 and R01 MH105561. Walter Offen is a former employee of AbbVie and is retired. All authors may own AbbVie stock. The authors would also like to thank an anonymous Associate Editor and anonymous referees for their many insightful comments and suggestions. The R code is available at \url{https://github.com/tian-yu-zhan/DNN_optimization}.
%
%
%
%
%
%
%
%
%
%

\bigskip

\bibliographystyle{Chicago}

\bibliography{ref}
\end{document}